\documentstyle[aps,epsfig,pra]{revtex}
\draft
\begin{document}
\def\la{\langle}
\def\ra{\rangle}
\def\om{\omega}
\def\Om{\Omega}
\def\vep{\varepsilon}
\def\wh{\widehat}
\newcommand{\beq}{\begin{equation}}
\newcommand{\eeq}{\end{equation}}
\newcommand{\beqa}{\begin{eqnarray}}
\newcommand{\eeqa}{\end{eqnarray}}
\newcommand{\intf}{\int_{-\infty}^\infty}
\newcommand{\into}{\int_0^\infty}
%

\begin{title}
{\Large\bf Comment on ``Measurement of time of arrival in quantum
mechanics''}
\end{title}
\author{\large A. D. Baute$^{1,2}$, I. L. Egusquiza$^1$ and J. G. Muga$^2$}
\address{
${}^1$ Fisika Teorikoaren Saila,
Euskal Herriko Unibertsitatea,
644 P.K., 48080 Bilbao, Spain\\
${}^2$ Departamento de Qu\'\i mica-F\'\i sica,
Universidad del Pa\'\i s Vasco, Apdo. 644, Bilbao, Spain
}
\date{\today}
\maketitle

\begin{abstract}
The analysis of the model quantum clocks proposed by Aharonov et al.
\cite{AOPRU} requires considering evanescent components, previously
ignored. We also clarify the meaning of the operational time of
arrival distribution which had been investigated.

\end{abstract}

\pacs{PACS: 03.65.-w
\hfill  EHU-FT/0010}

The concept of time has a very problematic status in the development of
quantum theory.  This has led several researchers to investigate and
propose quantum clocks, i.  e.  quantum devices that in some way will
capture and reflect a given aspect of time.  A recent such set of
proposals for quantum clocks, in the specific case of measurement of
times of arrival, has been put forward by Aharonov et al.
\cite{AOPRU}.  We shall concentrate here on the first quantum clock
proposed in that paper, which is coupled to the otherwise free system
whose times of arrival one desires to measure in such a way that the
total Hamiltonian (system plus clock) is
\begin{equation}
	H=\frac{1}{2m}{\bf P}_{x}^2 + \theta(-{\bf x}){\bf P}_{y}\,.
	\label{totham}
\end{equation}
The system variable is $x$, while the clock corresponds to the time
variable $y$. This is in fact a cyclic variable, which entails the
conservation of the energy ${\bf P}_{y}$. Let us then consider ${\bf P}_{y}$
restricted to a  given value
$p$, in the classical case.

The motion of the particle (the system) is not fully free:
moving from left to right, it
runs into a step potential at $x=0$, either downwards if $p>0$, or
upwards for negative $p$.
For $p>0$, the classical $y$
variable encodes how much time elapses since the particle is
released and the clock is started, to the instant the particle crosses $x=0$.  
For 
negative barrier height $p$, the classical particle would be reflected
if its energy were not big enough to overcome the step, and the clock
variable $y$ would keep running after the particle is reflected.  This
suggests using predominantly positive $p$ values in the analysis, even
though quantum mechanically there is reflection even for downward
steps.

In the quantum case, restricting ourselves to an eigenspace of ${\bf
P}_{y}$ with eigenvalue $p$, either positive or negative, the
(generalized) eigenstates of the (restricted) Hamiltonian are i)
scattering states, with degeneracy two; and ii) ``evanescent'' states,
whose eigenvalue is not degenerate.  On choosing an adequate
orthogonal basis of these scattering and evanescent states, one can
simply write the time evolution of any given state.  However, for
states with support in the $x$ space restricted to one side ($x<0$, in
particular), there is a compact alternative expression in terms of an
integral over a path in complex momentum space which provides us with
the whole spacetime dependence of the state \cite{HWZ,BEM00b}.

More explicitly, consider an initial wavefunction in $(x,y)$ space,
assumed to be factorized, $\psi(x,y,0)=\psi_{1}(x)\psi_{2}(y)$, such
that $\psi_{1}(x)$ has no support on positive $x$, and compute its
Fourier components, that is to say
\[\psi(x,y,0)=\frac{1}{2\pi\hbar}\int_{-\infty}^{+\infty}dk\,
\int_{-\infty}^{+\infty}dp\,e^{ikx/\hbar}e^{ipy/\hbar}g(k) f(p)\,.\]
It is then the case that for any time $t$, the state evolved with the
total Hamiltonian $H$, defined in Eq.  (\ref{totham}), can be written
as
\begin{equation}
	\psi(x,y,t)=\int_{-\infty}^{+\infty}dp \int_{\Gamma(p)} dk\,
	f(p) g(k) \phi_{kp}(x,y,t),
	\label{gammaevol}
\end{equation}
where
\begin{equation}
	\phi_{kp}(x,y,t)=\frac{1}{2\pi\hbar}\times\cases{
	\left(e^{ikx/\hbar}+\frac{k-q}{k+q}e^{-ikx/\hbar}\right) e^{ipy/\hbar}
	e^{-i\left(p+k^{2}/2m\right)t/\hbar}\,,& $\,\, x\leq0$\cr
	\frac{2k}{k+q}e^{iqx/\hbar}e^{ipy/\hbar}
	e^{-i\left(p+k^{2}/2m\right)t/\hbar}\,,& $\,\, x\geq0$\cr}
	\label{phikp}
\end{equation}
$q=\sqrt{k^{2}+2mp}$, defined with a branch cut in the $k$ plane,
which, for positive $p$, goes from $-i\sqrt{2mp}$
to $i\sqrt{2mp}$, and $\Gamma(p)$ is a path in the complex $k$
plane from real negative to real positive infinity which goes
\emph{above} the branch cut.

Thus, in order to write the general solution for Schr\"odinger's
equation with the Hamiltonian $H$ and initial support at $x<0$, it is
imperative to take into account the contribution of the branch cut in
Eq.  (\ref{gammaevol}), which was omitted in  \cite{AOPRU}.
It should be noted that supposing that $g(k)$ is zero
along the branch cuts would in turn bring in the problem that $g(k)$
would be forced to be either zero everywhere or non analytic; however,
if the initial wavefunction is normalizable and has initial support
at $x<0$, its Fourier transform $g(k)$ must be analytic in the upper
half plane. It follows that we cannot consistently assume both initial
localization and that $g(k)$ is zero along the branch cuts, and the
contribution of the branch cuts is of necessity present.  Quite another
issue is whether the contribution of the branch cut can be neglected
with regard to the physics that we want to describe.  If the state
were initially peaked at high energies, it would present negligible
overlap with evanescent states, and their contribution to the
posterior evolution of the state would remain ignorable.  However, we
should point out that generically the evanescent component cannot be
ignored, even in the asymptotic limit $t\to\infty$.  As a first
example, to be supplemented later graphically, consider the total
probability that a particle, which starts at $t=0$ from the left hand
side, $x<0$, has to be found at positive $x$ for large times, after
colliding with a downward step barrier of height $p$.  This
(transmission) probability tends to
\begin{equation}
	P_{T}=\int_{0}^\infty dq\,\left|\frac{2q}{q+k}g(k)\right|^2,
	\label{probtrans}
\end{equation}
where $q$ and $k$ are related as before, and the evanescent component
provides the lower part of the integration interval, from $0$ to
$\sqrt{2mp}$ (this expression can be obtained by using a variant of
Riemann-Lebesgue's lemma, in a similar way to the computation
performed by Allcock in the free case \cite{Allcock69}).
Note that there is contribution to (\ref{probtrans})
from positive and imaginary $k$, but 
not from negative values of $k$. 

\begin{figure}[h]
\epsfysize=8cm
\centerline{\epsfbox{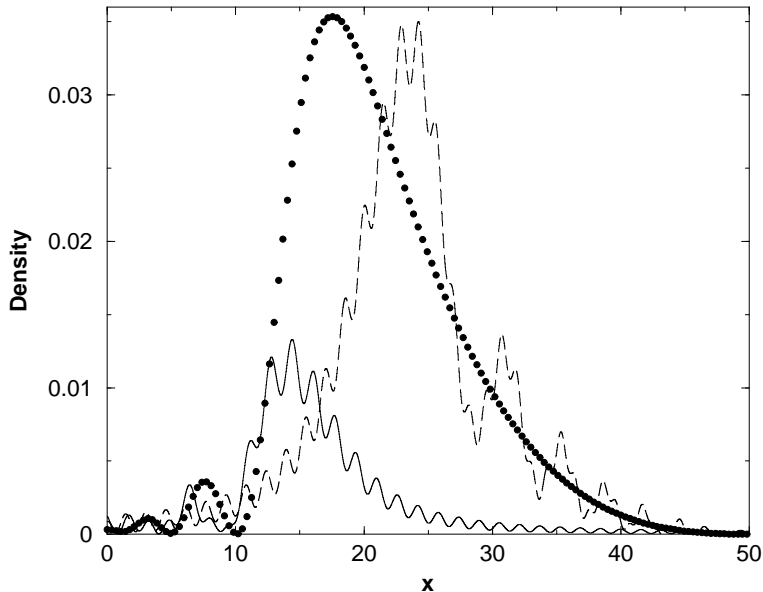}}
\caption{Collision with a downward step potential: The continuous line
portrays the modulus squared of the contribution of evanescent waves
to the wavefunction; the dashed line is assigned to the modulus
squared for scattering waves; the dotted line corresponds to the total
probability density.  All quantities are displayed in atomic units.  A
time $t=10$ a.u.  has elapsed.  Step height $p=2$ a.u.  . The initial
state is a truncated sine (see text), with $a=-2.01$ a.u.  and
$b=-0.01$ a.u.  .}\label{figone}
\end{figure}

In order to illustrate graphically the relevance of evanescent waves,
let us first examine figure \ref{figone}, where we depict the
probability density for positive $x$ at a given instant, as well as
the modulus square of the contribution to the wavefunction of
evanescent components (corresponding to imaginary $k$ along the branch cut
in (\ref{phikp})),
and the modulus square of the contribution of
scattering (real $k$) components, in the case of a step potential (fixed $p$);
that is to say, we are computing the result of a collision of a
wavefunction with a downward step potential.  The initial state is a
truncated sine function, i.e.,
$\psi_{0}(x)=(2/(b-a))^{1/2}\sin(\pi(x-a)/(b-a))
\left(\theta(x-a)-\theta(x-b)\right)$, where $\theta(x)$ is the step
function and $0>b>a$.  It is apparent that evanescent waves are
numerically important, and even more so the interference term between
the evanescent and scattering components.

\begin{figure}[h]
\epsfysize=8cm
\centerline{\epsfbox{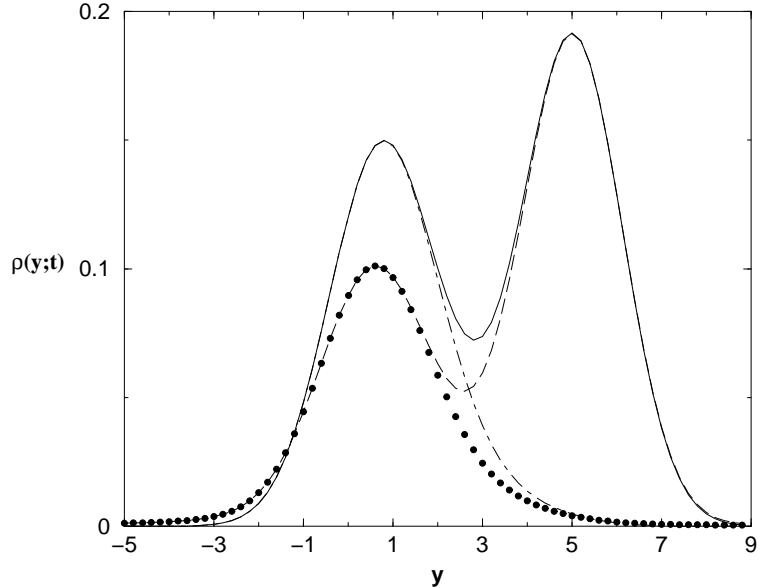}}
\caption{The conditional distribution (\ref{unilateral}) without
evanescent components is represented with a dotted line; with
evanescent components with the dashed-dotted line. The dashed line
corresponds to the unconditional distribution (\ref{bilateral})
without evanescent components; the continuous line depicts the
full unconditional distribution, with the evanescent components taken
into account. The initial state is a truncated sine
wave for the particle, as before, in $x$ space, and a minimum-uncertainty-product Gaussian 
for the clock, 
with central $p$ being $2$ a.u., $\la y\ra=0$,
and $\Delta y=1.1$ a.u.}
\label{figdos}
\end{figure}

{}From classical considerations, Aharonov et al. were led to
consider as an operational candidate for the distribution of times of
arrival of the $x$ particle the following expression:
\begin{equation}
	\rho_{c}(y,t)=\int_{0}^{\infty}dx\,|\psi(x,y,t)|^{2}.
	\label{unilateral}
\end{equation}
We would suggest that this quantity be better understood as an
operational distribution of \emph{dwell} times \emph{conditional} on
the particle being found at positive $x$. The corresponding
\emph{unconditional} operational
distribution would be given by
\begin{equation}
	\rho_{u}(y,t)=\int_{-\infty}^{+\infty}dx\,|\psi(x,y,t)|^{2}.
	\label{bilateral}
\end{equation}
The definition of the unconditional distribution makes apparent its
normalization to unity. However, if the evanescent components were not
included, there would be a probability deficit. Carrying back this
argument to the conditional distribution (\ref{unilateral}), which
need not  be normalized, one sees
that missing out the evanescent components leads to a probability
deficit. This is apparent in Fig. (\ref{figdos}).

Another effect of considering or not considering the
evanescent components is that the peak of arrivals is slightly
shifted towards higher values of $y$. This can be understood as due
to the fact that the evanescent components are, in a way, ``slower''
than those going over the step, thus remaining longer in $x<0$. This
also brings down the tail in negative $y$ coordinates.

The additional conceptual distinction between arrival and
dwell time distributions is due to the fact that, even in a classical
picture, particles that arrived at $x=0$ but did not cross this
point, being reflected, would force the clock dial $y$ to keep on
moving. Similarly, particles that had crossed $x=0$ and were later
reflected back would have the corresponding dial stop for a while and
then resume its ticking. In this way both $\rho_{c}$ and $\rho_{u}$
are (conditional and unconditional) distributions of dwell-times of the 
particle in the 
left half line.

Another way of understanding the need of reservations with respect to
the interpretation of $\rho_{c}$ as a time-of-arrival distribution 
comes about because of
Allcock's analysis of the possibility of ideal distribution
of times of arrival \cite{Allcock69}.  He identified the final
($t\to\infty$) probability of finding the particle in the half-space
of positive $x$ with the final probability of having arrived at $x=0$,
for a particle initially confined at negative $x$.  He concluded from
this identification that an ideal time-of-arrival probability
distribution could not exist; however, it is by now well known that an
ideal time-of-arrival distribution does indeed exist for the free
particle, namely, Kijowski's distribution \cite{Kijowski74,EM}.  As
has been discussed elsewhere \cite{ML,BEM00a}, the flaw in Allcock's 
argument was to 
ignore that negative momenta components (in the free particle case
and in the case at hand too) would contribute to the arrival
probability in a transient regime, thus invalidating the stated identification.
Similarly, $\rho_{c}(t\to\infty)$ only contains asymptotic information on dwell 
times, but not detailed information on the transients, so it   
cannot generically be interpreted 
as a distribution of times of arrival.

\begin{figure}[h]
\epsfysize=8cm
\centerline{\epsfbox{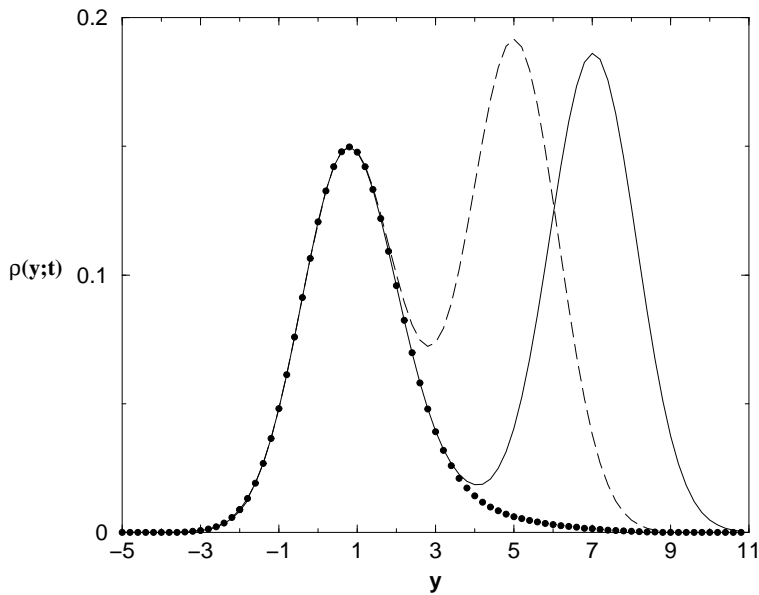}}
\caption{The full conditional distributions at times
$t=5$ a.u. and  $t=7$ a.u., for the same initial wavefunction as in
Fig. (\ref{figdos}), are shown with the dotted line.
The full unconditional distribution at $t=5$ a.u. is depicted with a
dashed line, while the continuous line represents the full
unconditional distribution at $t=7$ a.u. .}\label{figtres}
\end{figure}

The major difference between the conditional and unconditional
distributions, when enough time has passed for the situation to be
considered asymptotic, is the presence in the unconditional
distribution of a second peak, associated to those components of the
wavefunction that are still in $x<0$ and will in fact remain there.
This second peak is present no matter whether we do or do not take into
account the evanescent components, and, in fact, does not get any
contribution from them, as can be seen in Fig. (\ref{figdos}). The
evolution of the second peak, moving forward in $y$ space as time
goes by, is represented in Fig. (\ref{figtres}).

At any rate, it should be noticed that the numerical change of the
conditional distribution when the evanescent
components are taken into account, in the
asymptotic regime, is mostly a scale change, thus rendering the
difference irrelevant for the analysis carried out in \cite{AOPRU} with
regard to uncertainties in the position of the clock dial.
In particular, the lower bound suggested in  
\cite{AOPRU} for the product of the average energy of the particle and the
uncertainty in the 
time-of-arrival has been demonstrated for Kijowski's distribution 
without recourse to any coupling with clock variables \cite{BSPME00}. 

\acknowledgements

We acknowledge support by Ministerio de Educaci\'on y Cultura
(AEN99-0315), The University of the Basque Country (grant UPV
063.310-EB187/98), and the Basque Government (PI-1999-28).  A. D.
Baute acknowledges an FPI fellowship by Ministerio de Educaci\'on y
Cultura.

\end{document}